\documentclass[conference,compsocconf,letterpaper]{IEEEtran}
\makeatletter
\def\ps@headings{%
\def\@oddhead{\mbox{}\scriptsize\rightmark \hfil \thepage}%
\def\@evenhead{\scriptsize\thepage \hfil \leftmark\mbox{}}%
\def\@oddfoot{}%
\def\@evenfoot{}}
\makeatother
\usepackage{cite}
\usepackage{graphicx}
\usepackage{epstopdf}
\usepackage{amsthm}
\usepackage{amsmath}
\usepackage{xcolor}
\usepackage{amsfonts}
\usepackage{nicefrac}
\usepackage{amssymb,bm,upgreek}
\usepackage{algorithm}
\usepackage{nicefrac}
\usepackage{algpseudocode}
\usepackage{flushend}
\IEEEoverridecommandlockouts
\theoremstyle{remark}
\newtheorem{theorem}{\bf{Theorem}}
\newtheorem{proposition}{\bf{Proposition}}

\begin{document}
\title{Downlink Power Control in Massive MIMO Networks with Distributed Antenna Arrays}
\author{\IEEEauthorblockN{Noman Akbar$^{*}$, Emil Bj\"ornson$^{\dagger}$, Erik G. Larsson$^{\dagger}$, and Nan Yang$^{*}$}
\IEEEauthorblockA{$^{*}$Research School of Engineering, Australian National University, Acton, ACT, 2601,
Australia}\IEEEauthorblockA{$^{\dagger}$Department of Electrical Engineering (ISY), Link\"oping University, Sweden} Email: noman.akbar@anu.edu.au, emil.bjornson@liu.se, erik.g.larsson@liu.se, nan.yang@anu.edu.au \thanks{This research work was performed when the first author visited the Division of Communication Systems, Department of Electrical Engineering (ISY) in Link\"oping University, Sweden.}}

\markboth{Submitted to IEEE Globecom 2017}{Akbar \MakeLowercase{\textit{et
al.}}: Downlink Power Control in Distributed Antenna Array Massive MIMO Networks}

\maketitle

\begin{abstract}
In this paper, we investigate downlink power control in massive multiple-input multiple-output (MIMO) networks with distributed antenna arrays. The base station (BS) in each cell consists of multiple antenna arrays, which are deployed in arbitrary locations within the cell. Due to the spatial separation between antenna arrays, the large-scale propagation effect is different from a user to different antenna arrays in a cell, which makes power control a challenging problem as compared to conventional massive MIMO. We assume that the BS in each cell obtains the channel estimates via uplink pilots. Based on the channel estimates, the BSs perform maximum ratio transmission for the downlink. We then derive a closed-form spectral efficiency (SE) expression, where the channels are subject to correlated fading. Utilizing the derived expression, we propose a max-min power control algorithm to ensure that each user in the network receives a uniform quality of service. Numerical results demonstrate that, for the network considered in this work, optimizing for max-min SE through the max-min power control improves the sum SE of the network as compared to the equal power allocation.
\end{abstract}

\section{Introduction}
Massive multiple-input multiple-output (MIMO) is widely acknowledged as a key enabling technology for the next generation mobile communication networks. Massive MIMO offers an increased spectral and energy efficiency as compared to regular MIMO  \cite{Marzetta2010,Yang2015}. In massive MIMO, the base stations (BSs) are equipped with a very large number of antennas and serve many users simultaneously. The BSs' antenna arrays can be deployed in either a co-located or a distributed manner. In co-located deployment, the BS consists of a single antenna array, where all the antenna elements are in close proximity to each other. Differently, in distributed deployment, the antenna arrays belonging to a BS are not necessarily deployed in close proximity to each other \cite{Ngo2017,Zhou2003}. As such, the antenna arrays can be placed at arbitrary locations within a cell. We refer to this as distributed antenna array (DAA) massive MIMO.

Most existing works in the massive MIMO literature assume a co-located antenna array deployment \cite{Liu2014,Emil2016b,Akbar2016a,Akbar16b}. However, DAA massive MIMO can offer a number of advantages over the co-located antenna array massive MIMO. For example, it has the potential to improve the capacity and the coverage, as the users are closer to the antenna arrays \cite{Choi2007,Ngo2017,Emil2015,Castanheira2010}. Additionally, the DAA deployment offers more resilience to shadow fading than the co-located deployment at the cost of an increased back-haul communication requirements \cite{Ngo2017}. In this paper, we assume that the antenna arrays are deployed in a distributed manner at arbitrary locations. Each antenna array is connected to a cell processing unit (CPU) through a back-haul link, which enables coherent processing of the signal transmitted from the arrays. Each cell in the network can be considered as a ``Cell-Free'' massive MIMO \cite{Nayebi2015,Ngo2017}, where the access points have multiple antenna elements. We highlight that the system model considered in this paper is a generalized model, which encompasses the ``Cell-Free'' massive MIMO as a special case.

Power control in massive MIMO is a pivotal technique to achieve a uniform quality of service for every user throughout the network. We highlight that the power control is performed to utilize the available power in an efficient manner. Although power control in massive MIMO is a well investigated topic \cite{Li2016,Cheng2017,Marzetta2016}, the optimal power control in multi-cell DAA massive MIMO is a new problem. Specifically, this power control problem is challenging because the large-scale propagation effect varies from a user to different antenna arrays in a cell. The optimal power allocation problem in ``Cell-Free'' massive MIMO was investigated in \cite{Nayebi2015,Ngo2017}, where each access point was equipped with a single antenna element. Differently, in this work we consider that each array has multiple antenna elements and the network has a conventional cellular structure.

The novel contributions of our paper are:
\begin{enumerate}
\item
We derive a closed-form expression for the downlink signal-to-interference-plus-noise ratio (SINR) in DAA massive MIMO under the assumption of correlated channel fading. The expression is valid for an arbitrary number of DAAs.
\item
We investigate optimal downlink power allocation and equal power allocation in a DAA massive MIMO network. To this end, we formulate the downlink power control problem as a max-min optimization problem.
\item
We present numerical results for max-min power allocation and equal power allocation.
\end{enumerate}
\textit{Notations:} We denote vectors and matrices by lower-case boldface symbols and upper-case boldface symbols, respectively. $\mathbb{E}[\cdot]$ denotes the expectation, $\|\cdot\|$, denotes the $\textit{l}_2$ norm, $\textrm{tr}(\cdot)$ denotes the matrix trace, $(\cdot)^{\textrm{H}}$ denotes the Hermitian transpose, $(\cdot)^{T}$ denotes the matrix transpose, and $\mathbf{I}_M$ denotes an $M\times M$ identity matrix.

\section{Distributed Antenna Array Massive MIMO Network}
\begin{figure}[!t]
\centering
\includegraphics[width=21pc]{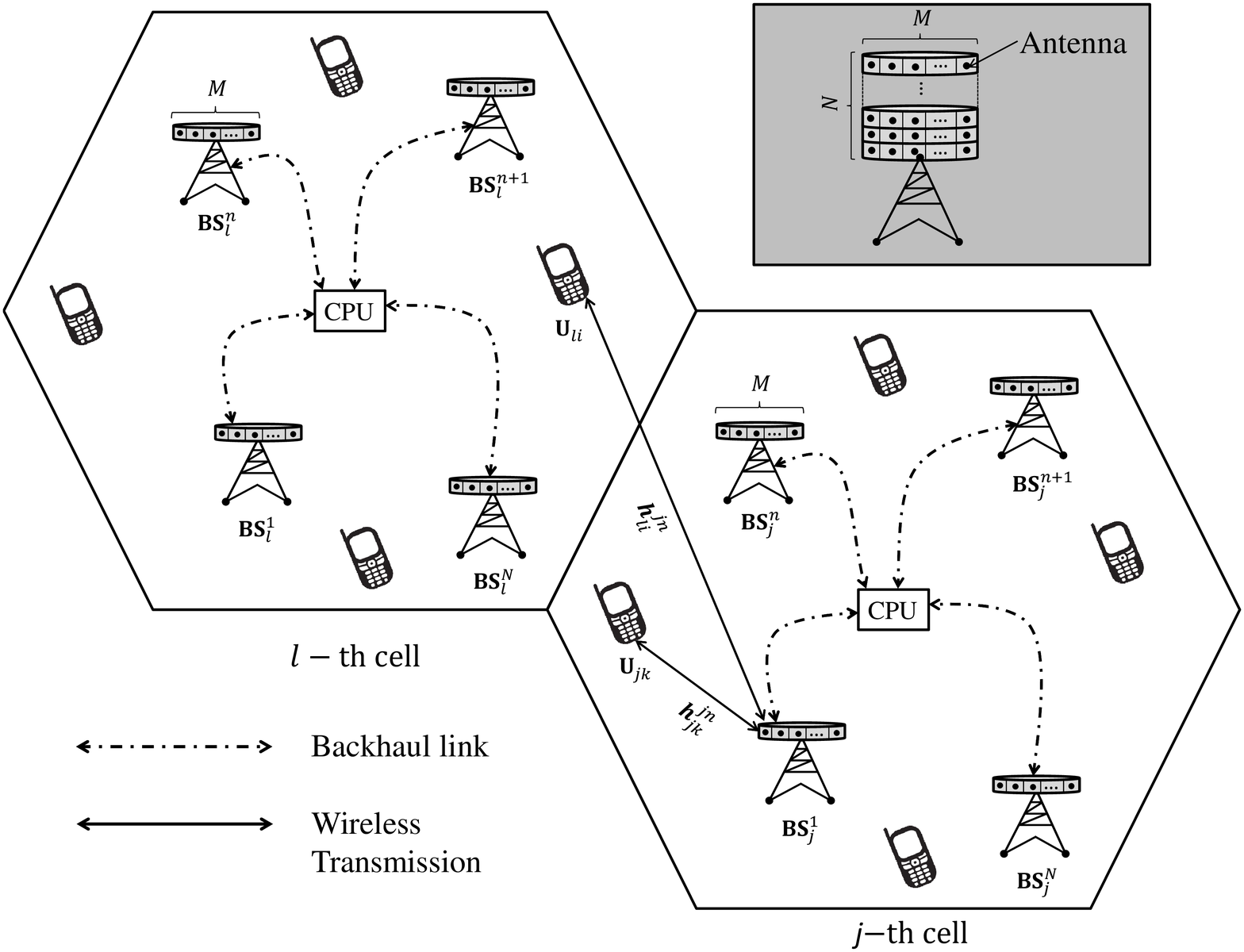}
\caption{Illustration of multi-cell DAA massive MIMO with $N$ arrays in each cells. Each antenna array is equipped with $M$ antenna elements. The shaded box in top right corner depicts a conventional co-located massive MIMO BS.}
\label{system_model}
\end{figure}

We consider a multi-cell network consisting of $L$ cells and $K$ single--antenna users in each cell. We assume that each cell has DAA BSs. As such, each BS in a cell is equipped with $N$ DAAs deployed at arbitrary locations, where each antenna array has $M$ antenna elements as depicted in Fig.~\ref{system_model}. We denote the \textit{k}-th user in the \textit{j}-th cell by $\textrm{U}_{jk}$, where $j\in\{1,\cdots,L\}$ and $k\in\{1,\cdots,K\}$. Moreover, we denote the BS in the \textit{j}-th cell by $\textrm{BS}_{j}$ and the \textit{n}-th antenna array belonging to $\textrm{BS}_{j}$ by $\textrm{BS}_{j}^n$. Furthermore, we denote the uplink channel between $\textrm{U}_{li}$ and $\textrm{BS}_{j}^n$ by $\mathbf{h}_{li}^{jn}$, where $l\in\{1,\cdots,L\}$, $i\in\{1,\cdots,K\}$, and $n\in\{1,\cdots,N\}$. In this paper, we assume that the channels follow a correlated Rayleigh fading distribution $\mathbf{h}_{li}^{jn} \sim \mathcal{CN}(\mathbf{0},\mathbf{R}_{li}^{jn})$, where $\mathbf{R}_{li}^{jn}$ is the channel covariance matrix that captures various channel properties such as average path-loss and spatial correlation.

We assume that the network operates in time division duplex (TDD) mode \cite{Marzetta2010,Akbar2016a}. As such, the uplink and downlink channels are assumed to be the same during a coherence time-frequency interval and independent between different intervals. The transmission consists of uplink pilots for channel estimation followed by downlink data transmission. We assume that the entire transmission is carried out inside one coherence time-frequency interval.

\subsection{Uplink Channel Estimation}
\begin{figure*}[!t]
\begin{align} \label{SINR_chan}
\gamma_{jk}=\frac{\left| \textstyle{\sum_{n=1}^N} \nu_{jk}^n \mathbb{E}\left[(\mathbf{h}_{jk}^{jn})^\textrm{H} \mathbf{{a}}_{jk}^{n} \right]\right|^2}{\textstyle{\sum_{l=1}^L\sum_{i=1}^K} \mathbb{E}\left[\left|\sum_{n=1}^N\nu_{li}^n(\mathbf{h}_{jk}^{ln})^\textrm{H} \mathbf{{a}}_{li}^{n}\right|^2\right] - \left|\textstyle{\sum_{n=1}^N} \nu_{jk}^n\mathbb{E}\left[(\mathbf{h}_{jk}^{jn})^\textrm{H} \mathbf{{a}}_{jk}^{n}\right]\right|^2 + \sigma_{n}^2}. \tag{11}
\end{align}
\begin{align} \label{SINR}
{\gamma}_{jk}=\frac{\left|\textstyle{\sum_{n=1}^N}\textrm{tr}\left(\nu_{jk}^n\mathbf{W}_{jk}^{n}\mathbf{R}_{jk}^{jn}\right)\right|^2} {\textstyle{\sum_{l=1}^L \sum_{i=1}^K\sum_{n=1}^N} \textrm{tr}\left((\nu_{li}^n)^2\mathbf{W}_{li}^n \mathbf{Q}_{li}^n(\mathbf{W}_{li}^n)^\textrm{H}\mathbf{R}_{jk}^{ln}\right) + \textstyle{\sum_{l=1,l\neq j}^L} \left|\sum_{n=1}^N\textrm{tr}\left(\nu_{lk}^n\mathbf{W}_{lk}^n\mathbf{R}_{jk}^{ln}\right)\right|^2 + \sigma_{n}^2}. \tag{12}
\end{align}
\hrulefill
\vspace*{4pt}
\end{figure*}
During the uplink channel estimation phase, all users in the network send their pre-assigned pilot sequences to the same-cell BSs. Specifically, $\mathbf{\boldsymbol{\phi}}_{jk}$ is the pilot sequence associated with $\textrm{U}_{jk}$ such that $\|\mathbf{\boldsymbol{\phi}}_{jk}\|^2=1$. We assume that all pilot sequences have length $\tau_p$. In this work we assume that $\tau_p=K$. As such, each user in a cell is assigned an orthogonal pilot sequence. Furthermore, the same set of pilot sequences are repeated in each cell across the network. Notably, the results can be easily generalized to other cases for pilot assignment. Accordingly, the uplink pilot transmission received at the \textit{n}-th sub-array of $\textrm{BS}_{j}$, i.e., $\textrm{BS}_{j}^n$, is given as
\begin{align} \label{uplink_trans}
\mathbf{Y}^{jn} &= \sum\limits_{l=1}^L\sum\limits_{i=1}^K\mathbf{h}_{li}^{jn}\mathbf{\boldsymbol{\phi}}_{li}^\textrm{H} + \frac{1}{\sqrt{\rho_{\textrm{tr}}}}\mathbf{N}_{j}^n,
\end{align}
where $\mathbf{N}_{j}^n\in \mathbf{\mathbb{C}}^{M \times \tau}$ represents the additive white Gaussian noise (AWGN) at $\textrm{BS}_{j}^n$, and $\rho_{\textrm{tr}}=\rho_p\tau_p$ is the normalized pilot power per user. The sub-array $\textrm{BS}_{j}^n$ correlates \eqref{uplink_trans} with the known pilot sequence to obtain
\begin{align} \label{uplink_trans1}
\mathbf{y}_{jk}^{jn} &= \left(\sum\limits_{l=1}^L\sum\limits_{i=1}^K\mathbf{h}_{li}^{jn}\mathbf{\boldsymbol{\phi}}_{li}^\textrm{H} + \frac{1}{\sqrt{\rho_{\textrm{tr}}}}\mathbf{N}_{j}^n\right)\mathbf{\boldsymbol{\phi}}_{jk}.
\end{align}
Assuming that the \textit{k}-th user in each cell is assigned the same pilot sequence, we have $\mathbf{\boldsymbol{\phi}}_{li}^\textrm{H}\mathbf{\boldsymbol{\phi}}_{jk} = 1$ when $i=k$ and $\mathbf{\boldsymbol{\phi}}_{li}^\textrm{H}\mathbf{\boldsymbol{\phi}}_{jk}= 0$ when $i\neq k$. Thus, we simplify \eqref{uplink_trans1} as
\begin{align} \label{uplink_trans2}
\mathbf{y}_{jk}^{jn} &= \mathbf{h}_{jk}^{jn} + \sum\limits_{l=1,l\neq j}^L\mathbf{h}_{lk}^{jn} + \frac{1}{\sqrt{\rho_{\textrm{tr}}}}\mathbf{N}_{j}^n\mathbf{\boldsymbol{\phi}}_{jk}.
\end{align}
From \eqref{uplink_trans2}, we obtain the MMSE estimate of $\mathbf{h}_{jk}^{jn}$ as \cite{Emil2016a}
\begin{align} \label{mmse_est}
\mathbf{\widehat{h}}_{jk}^{jn} &= \mathbf{W}_{jk}^{jn}\mathbf{y}_{jk}^{jn},
\end{align}
where $\mathbf{W}_{jk}^{jn}=\mathbf{R}_{jk}^{jn}(\mathbf{Q}_{jk}^{jn})^{-1}$, $\mathbf{R}_{jk}^{jn}=\mathbb{E}\left[\mathbf{h}_{jk}^{jn}(\mathbf{h}_{jk}^{jn})^\textrm{H}\right]$, and
\begin{align}\label{Q_val}
  \mathbf{Q}_{jk}^{jn}=\mathbb{E}\left[\mathbf{y}_{jk}^{jn}(\mathbf{y}_{jk}^{jn})^\textrm{H}\right] = \sum\limits_{l=1}^L\mathbf{R}_{lk}^{jn} + \frac{1}{\rho_{\textrm{tr}}}\mathbf{I}_M.
\end{align}
Throughout this paper, we assume that the covariance matrices $\mathbf{R}_{jk}^{jn}$ and $\mathbf{Q}_{jk}^n$ are known to the BSs.

\subsection{Downlink Data Transmission}
During the downlink data transmission phase, $\textrm{BS}_{j}$ transmits data symbols to each user in the \textit{j}-th cell. The channel estimates obtained through the uplink channel estimation are also utilized for downlink transmission under the consideration of the TDD mode. Accordingly, the symbol transmitted by $\textrm{BS}_j$ for the $K$ same-cell users is represented as
\begin{align}\label{symbol_n}
  x_{j} &= \sum_{i=1}^{K}\sum_{n=1}^{N} \nu_{ji}^n \mathbf{{a}}_{ji}^{n}q_{ji},
\end{align}
where $\nu_{ji}^n \geq 0$ is the real-valued downlink power control coefficient for $\textrm{U}_{ji}$ at $\textrm{BS}_{j}^n$, $\mathbf{{a}}_{ji}^{n}$ is the precoding vector for $\textrm{U}_{ji}$ at $\textrm{BS}_{j}^n$, $q_{ji}$ is the data symbol intended for $\textrm{U}_{ji}$, and ${q}_{ji} \sim \mathcal{CN}\left(0,1\right)$. We assume that the downlink power control coefficients are chosen to satisfy $\mathbb{E}\left[|x_{j}|^2\right] \leq 1$. This power constraint can be re-written as
\begin{align}\label{pow_constaint}
\sum_{i=1}^K \sum_{n=1}^{N} ({\nu_{ji}^n})^2\mathbb{E}\left[\|\mathbf{{a}}_{ji}^{n}\|^2\right] &\leq 1,~\forall~j.
\end{align}

The constraint in \eqref{pow_constaint} represents the total transmit power constraint in cell $j$, which is normalized such that the maximum power is 1. From \eqref{symbol_n}, the downlink transmission received at $\textrm{U}_{jk}$ is
\begin{align}\label{pow_constaint1}
r_{jk} &=\sum_{l=1}^L\sum_{i=1}^K\sum_{n=1}^N \nu_{li}^n (\mathbf{h}_{jk}^{ln})^\textrm{H} \mathbf{{a}}_{li}^{n}q_{li}+ n_{jk},
\end{align}
where $n_{jk}$ is the AWGN at $\textrm{U}_{jk}$. We assume that the users do not have knowledge about the instantaneous channel and only know the channel statistics \cite{Akbar2016a}. Accordingly, the downlink signal received at $\textrm{U}_{jk}$ is represented as
\begin{align} \label{received1}
r_{jk} &=\sum\limits_{n=1}^N \nu_{jk}^n \mathbb{E}\left[(\mathbf{h}_{jk}^{jn})^\textrm{H} \mathbf{{a}}_{jk}^{n}\right]q_{jk} + \underbrace{\textstyle{\sum_{l,i,n}^{L,K,N}}}_{(l,i) \neq (j,k)}\nu_{li}^n(\mathbf{h}_{jk}^{ln})^\textrm{H} \mathbf{{a}}_{li}^{n}q_{li} \nonumber \\ & +\sum\limits_{n=1}^N \nu_{jk}^n\bigg((\mathbf{h}_{jk}^{jn})^\textrm{H}\mathbf{{a}}_{jk}^{n} - \mathbb{E}\left[(\mathbf{h}_{jk}^{jn})^\textrm{H}\mathbf{{a}}_{jk}^{n}\right]\bigg)q_{jk} + n_{jk},
\end{align}
\normalsize
We note that the first term in \eqref{received1} represents $N$ superimposed copies of the symbol $q_{jk}$ received from the different arrays in cell $j$. We highlight that \eqref{received1} is a generalized expression for received signal at $\textrm{U}_{jk}$, which is valid for an arbitrary number of DAAs in a cell.

\subsection{Achievable Downlink Sum Spectral Efficiency}

In this subsection, we derive a closed-form expression for the downlink spectral efficiency (SE). We then compute the sum SE and use it as a performance metric. We note that the last three terms in \eqref{received1} can be considered as the effective noise and are uncorrelated with the first term in \eqref{received1}. Accordingly, the downlink SE for $\textrm{U}_{jk}$ is given as
\begin{align}\label{SE_chan}
\textrm{SE}_{jk}&=\left(1-\frac{K}{\tau_c}\right) \log_{2}\left(1+\gamma_{jk}\right)& \textrm{b/s/Hz},
\end{align}
where $\tau_c$ is the channel coherence interval in number of samples and $\gamma_{jk}$ is the downlink effective SINR for $\textrm{U}_{jk}$ given by \eqref{SINR_chan} at the top of the page. The sum SE in a cell is the sum of SE of all the same-cell users. We next provide a closed-form expression for the SINR in the following theorem.
\begin{theorem} \label{theorem}
Assuming that the BSs perform maximum ratio transmission (MRT) in the downlink, i.e, $\mathbf{a}_{jk}^{n} = \mathbf{\widehat{h}}_{jk}^{jn}$, the closed-form expression for the downlink effective SINR at $\textrm{U}_{jk}$ is obtained as in \eqref{SINR} at the top of the page.
\end{theorem}
\begin{IEEEproof}
Please see Appendix \ref{SINR_proof}.
\end{IEEEproof}
The closed-form expression for the downlink SINR given in \eqref{SINR} can be re-written as
\begin{align}\label{SINR_reduced}
{\gamma}_{jk}=\frac{\left|\textstyle{\sum_{n=1}^N}\nu_{jk}^n\chi_{jk}^{n}\right|^2} {\textstyle{\sum_{l,i,n}^{L,K,N}}(\nu_{li}^n)^2\zeta_{jk}^{lin} + \textstyle{\sum_{l\neq j}^{L}}|\textstyle{\sum_{n=1}^N}\nu_{lk}^n\xi_{jk}^{ln}|^2 + \sigma_{n}^2},\setcounter{equation}{12}
\end{align}
where
  $\chi_{jk}^{n} = \textrm{tr}(\mathbf{W}_{jk}^{n}\mathbf{R}_{jk}^{jn})$,
  $\zeta_{jk}^{lin} = \textrm{tr}(\mathbf{W}_{li}^n \mathbf{Q}_{li}^n(\mathbf{W}_{li}^n)^\textrm{H}\mathbf{R}_{jk}^{ln})$, and
  $\xi_{jk}^{ln} = \textrm{tr}(\mathbf{W}_{lk}^n\mathbf{R}_{jk}^{ln})$.
\section{Downlink Power Control in Distributed Antenna Array Massive MIMO}
In this section, we formulate the downlink power control problem as a max-min optimization problem. Max-min power control maximizes the minimum SE for all the user in the network. As such, every user in the network receives a uniform quality of service. Max-min power control have previously been studied for conventional BSs \cite{Yang2014,Marzetta2016}. However, the application of max-min power control for DAA massive MIMO networks, where each array has multiple antenna elements, has not been investigated.

The goal of the max-min optimization problem is to maximize the minimum SE for all the users in the network. As such, we formulate the max-min optimization problem using \eqref{SINR_reduced} as
\begin{align} \label{opt_problem}
\begin{aligned}
& \underset{\{\nu_{li}^n\}}{\text{max}}~~~\underset{\forall~j,k}{\text{min}}
& & \frac{\left|\textstyle{\sum_{n=1}^N}\nu_{jk}^n\chi_{jk}^{n}\right|^2} {\textstyle{\sum_{l,i,n}^{L,K,N}}(\nu_{li}^n)^2\zeta_{jk}^{lin} + \textstyle{\sum_{l\neq j}^{L}}|\textstyle{\sum_{n=1}^N}\nu_{lk}^n\xi_{jk}^{ln}|^2 + \sigma_{n}^2}. \\
& \text{s. t.}
& &  \hspace{-0.4cm} \textrm{tr}(\mathbf{W}_{jk}^{n}\mathbf{R}_{jk}^{jn}) \leq \chi_{jk}^{n},\forall~n,\\
& & &\hspace{-0.4cm} \textrm{tr}(\mathbf{W}_{li}^n \mathbf{Q}_{li}^n\left(\mathbf{W}_{li}^n\right)^\textrm{H}\mathbf{R}_{jk}^{ln}) \leq \zeta_{jk}^{lin} ,\forall~l,i,n,\\
& & & \hspace{-0.4cm} \textrm{tr}(\mathbf{W}_{lk}^n\mathbf{R}_{jk}^{ln})\leq \xi_{jk}^{ln}, \forall~l,n,\\
& & & \hspace{-0.4cm} \textstyle{\sum_{i=1}^K\textstyle{\sum_{n=1}^N}(\nu_{li}^n)^2\textrm{tr}(\mathbf{W}_{li}^n \mathbf{Q}_{li}^n\left(\mathbf{W}_{li}^n\right)^\textrm{H}) \leq 1},\forall~l,n,\\
& & & \hspace{-0.4cm} \nu_{li}^n \geq 0, \;\forall~l,i,n,
\end{aligned}
\end{align}
where the constraint $\sum_{i=1}^K\sum_{n=1}^N(\nu_{li}^n)^2\textrm{tr}(\mathbf{W}_{li}^n \mathbf{Q}_{li}^n\left(\mathbf{W}_{li}^n\right)^\textrm{H}) \leq 1$ is obtained from \eqref{pow_constaint} under the assumption that MRT is used at the BSs. Assuming that the target SINR is ${\gamma}$, we rewrite the optimization problem given in \eqref{opt_problem} in the epigraph form as
\begin{align} \label{opt_problem1}
\begin{aligned}
& \underset{\{\nu_{li}^n\},{\gamma}}{\text{max}}
& & \gamma \\
& \text{s. t.}
& & \hspace{-0.65cm} \frac{\left|\textstyle{\sum_{n=1}^N}\nu_{jk}^n\chi_{jk}^{n}\right|^2} {\textstyle{\sum_{l,i,n}^{L,K,N}}(\nu_{li}^n)^2\zeta_{jk}^{lin} + \textstyle{\sum_{l\neq j}^{L}}|\textstyle{\sum_{n=1}^N}\nu_{lk}^n\xi_{jk}^{ln}|^2 + \sigma_{n}^2} \geq {\gamma},\;\forall\;j,k, \\
& &  &\hspace{-0.65cm}\textrm{tr}(\mathbf{W}_{jk}^{n}\mathbf{R}_{jk}^{jn}) \leq \chi_{jk}^{n}, \forall~n,\\
& & &\hspace{-0.65cm}\textrm{tr}(\mathbf{W}_{li}^n \mathbf{Q}_{li}^n\left(\mathbf{W}_{li}^n\right)^\textrm{H}\mathbf{R}_{jk}^{ln}) \leq \zeta_{jk}^{lin}, \forall~l,i,n,\\
& & & \hspace{-0.65cm}\textrm{tr}(\mathbf{W}_{lk}^n\mathbf{R}_{jk}^{ln})\leq \xi_{jk}^{ln}, \forall~l,n,\\
& & & \hspace{-0.65cm} \textstyle{\sum_{i=1}^K\textstyle{\sum_{n=1}^N}(\nu_{li}^n)^2\textrm{tr}(\mathbf{W}_{li}^n \mathbf{Q}_{li}^n\left(\mathbf{W}_{li}^n\right)^\textrm{H}) \leq 1},\forall~l,n,\\
& & &\hspace{-0.65cm} \nu_{li}^n \geq 0,\; \forall~l,i,n.
\end{aligned}
\end{align}

This can be solved as a quasi-convex problem. We next formulate a convex feasibility problem based on \eqref{opt_problem1}, which we use in a bisection algorithm \cite{Boyd2004} to search for the value of $\gamma\in[\gamma_{\textrm{min}},\gamma_{\textrm{max}}]$ that is the global optimum to \eqref{opt_problem1}, where ${\gamma}_{\textrm{min}}$ and ${\gamma}_{\textrm{max}}$ define the search range \cite{Ngo2017,Boyd2004}.
\begin{proposition} \label{prop_1}
The constraint set in the optimization problem \eqref{opt_problem1} is convex and the optimization problem is quasi-concave. Assuming that $\gamma$ is a constant, the optimization problem in \eqref{opt_problem1} is re-written as the convex feasibility problem
\begin{align} \label{opt_problem2}
\begin{aligned}
& \underset{\{\nu_{li}^n\}}{\text{max}}
& & 0 \\
& \text{s. t.}
& &   \|\mathbf{x}_{jk}\| \leq \frac{1}{\sqrt{\gamma}}{\left|\textstyle{\sum_{n=1}^N}\nu_{jk}^n\chi_{jk}^{n}\right|},\; \forall~j,k, \\
& &  &\textrm{tr}(\mathbf{W}_{jk}^{n}\mathbf{R}_{jk}^{jn}) \leq \chi_{jk}^{n}, \forall~n,\\
& & &\textrm{tr}(\mathbf{W}_{li}^n \mathbf{Q}_{li}^n\left(\mathbf{W}_{li}^n\right)^\textrm{H}\mathbf{R}_{jk}^{ln}) \leq \zeta_{jk}^{lin} ,\;\forall~l,i,n\\
& & &\textrm{tr}(\mathbf{W}_{lk}^n\mathbf{R}_{jk}^{ln})\leq \xi_{jk}^{ln},\;\forall~l,n\\
& & & \nu_{lk}^n\xi_{jk}^{ln} \leq \varrho_{jk}^{lin}, \forall~l,n,\\
& & & \textstyle{\sum_{i=1}^K\sum_{n=1}^N(\nu_{li}^n)^2\textrm{tr}(\mathbf{W}_{li}^n \mathbf{Q}_{li}^n\left(\mathbf{W}_{li}^n\right)^\textrm{H}) \leq 1},\; \forall~l,\\
& & &\nu_{li}^n \geq 0,\;\forall\;l,i,n,
\end{aligned}
\end{align}
where $\mathbf{x}_{jk} = [\mathbf{\tilde{x}}_{jk}~~\mathbf{\bar{x}}_{jk}~~\sqrt{\sigma_{n}^2}]^T$. We define $\mathbf{\tilde{x}}_{jk}$ and $\mathbf{\bar{x}}_{jk}$ as $\mathbf{\tilde{x}}_{jk}=[\mathbf{\tilde{x}}_{jk}^{11}\dotsc\mathbf{\tilde{x}}_{jk}^{li}\dotsc\mathbf{\tilde{x}}_{jk}^{LK}]$ and $\mathbf{\bar{x}}_{jk}=[\mathbf{\bar{x}}_{jk}^{11}\dotsc\mathbf{\bar{x}}_{jk}^{li}\dotsc\mathbf{\bar{x}}_{jk}^{LK}]$, respectively, where
\begin{align}\label{def_x11}
\mathbf{\tilde{x}}_{jk}^{li}=[(\nu_{li}^1)(\zeta_{jk}^{li1})^{\frac{1}{2}}\dotsc(\nu_{li}^N)(\zeta_{jk}^{liN})^{\frac{1}{2}}]
\end{align}
and
\begin{align}\label{def_x11}
\mathbf{\bar{x}}_{jk}^{li}=\begin{cases}
\varrho_{jk}^{lk1}+\varrho_{jk}^{lkn}+\dotsc+\varrho_{jk}^{lkN} & l\neq j,\\
{0}, & l=j.
\end{cases}
\end{align}
\end{proposition}
\begin{IEEEproof}
Please see Appendix \ref{SOCP_proof}.
\end{IEEEproof}

In each iteration of the bisection algorithm, we set $\bar{\gamma}=({\gamma}_{\textrm{min}}+{\gamma}_{\textrm{max}})/2$ and solve the feasibility problem \eqref{opt_problem2} by setting ${\gamma}=\bar{\gamma}$. If the problem is infeasible, we set ${\gamma}_{\textrm{max}}=\bar{\gamma}$ otherwise we set ${\gamma}_{\textrm{min}}=\bar{\gamma}$. The algorithm iteratively refines ${\gamma}_{\textrm{min}}$ and ${\gamma}_{\textrm{max}}$ and stops the search when ${\gamma}_{\textrm{max}}-{\gamma}_{\textrm{max}}<\varepsilon$, where $\varepsilon>0$ is the error tolerance. We highlight that the max-min power control in ``Cell Free'' massive MIMO \cite{Nayebi2015} is a special case of the power control problem considered in this paper, which can be obtained from \eqref{opt_problem2} when the network has one cell and each antenna array has one antenna element.

The max-min power control in \eqref{opt_problem1} maximizes the minimum SE. In this work, we evaluate sum SE to demonstrate the performance of the power control algorithms. We highlight that the SE is dependent on the effective SINR as given in \eqref{SE_chan}. Therefore, by evaluating sum SE, we also demonstrate the improvement in SE.
\subsection{Equal Power Allocation}
The baseline for comparison is equal power allocation. Here, the total available downlink transmit power is shared equally among all the users in a cell \cite{Yang2014}. As such, the downlink power control coefficients ${\nu_{li}^n}$ are equal for all the users. From \eqref{pow_constaint} and assuming that the full available power is used by the BSs during the downlink transmission, we obtain
\begin{align}\label{pow_constaint_equal1}
(\nu)^2\sum_{l=1}^L\sum_{i=1}^K \sum_{n=1}^{N} \mathbb{E}\left[\|\mathbf{{a}}_{li}^{n}\|^2\right] &= L.
\end{align}
Assuming that the BSs performs MRT, the power control coefficient is obtained as
\begin{align}\label{equal_pow}
  {\nu} &= \sqrt{\frac{L}{\sum_{l=1}^L\sum_{i=1}^K \sum_{n=1}^{N}\textrm{tr}\left(\mathbf{W}_{li}^n \mathbf{Q}_{li}^n\left(\mathbf{W}_{li}^n\right)^\textrm{H}\right)}}.
\end{align}

With equal power allocation, the power control coefficients $\nu$ remain the same regardless of the channel conditions. Accordingly, it is expected that equal power allocation does not give a higher sum SE as compared to the max-min power control. However, equal power allocation serves as an important benchmark for the network performance.

\section{Numerical Results}
\begin{figure}[!t]
\centering
\includegraphics[width=16pc]{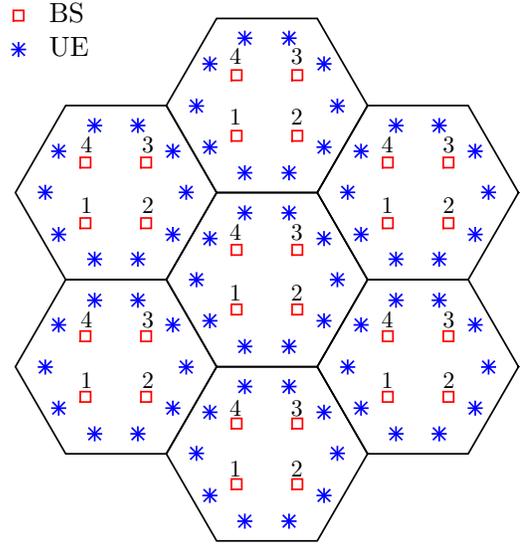}
\caption{DAA massive MIMO network with $N=4$ antenna arrays in each cells. Each antenna array is equipped with $M$ antennas.}
\label{simulated_nw}
\end{figure}
In this section, we compare the performance of the downlink power control schemes namely, max-min power control and equal power allocation. We consider a network with $L=7$ cells and DAAs, as depicted in Fig.~\ref{simulated_nw}. We assume that the DAAs are placed at 300\,m from the cell center. Additionally, we assume that each cell has $K=10$\,users and all users are 700\,m away from the center of the cell as shown in Fig.~\ref{simulated_nw}. For such a network configuration, the channel covariance matrices are computed using the one-ring model \cite{Adhikary2013}. We assume that $\tau_c=200$\,samples. These simulation parameters are kept the same throughout this section unless stated otherwise.

\begin{figure}[!t]
\centering
\includegraphics[width=21pc]{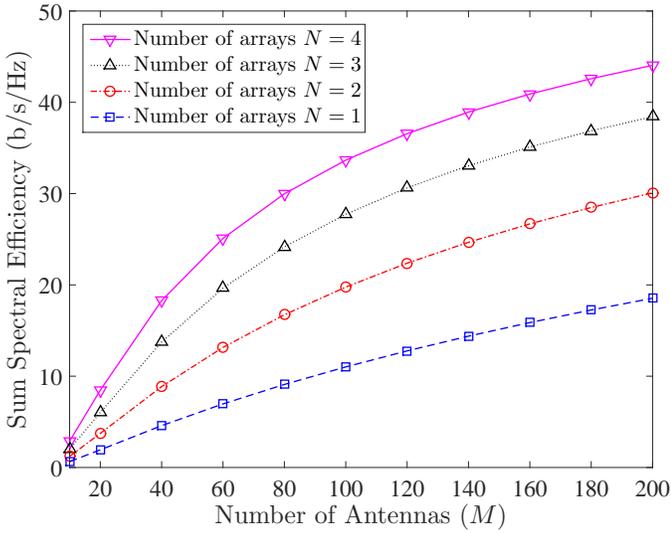}
\caption{The sum SE versus  the number of antennas per array for different $N$.}
\label{sub_arrays}
\end{figure}

We first examine the impact of increasing the number of arrays in a cell on the sum SE of the network. The advantages of increasing the number of antenna arrays are clearly observed from Fig.~\ref{sub_arrays}. The result is obtained using \eqref{SINR} for equal power allocation. In this simulation, for a given $N$, all DAAs marked $N$ or lower in Fig.~\ref{simulated_nw} are active. For example, when $N=3$, the antenna arrays marked $1$, $2$, and $3$ are active. We note that additional antenna arrays in the network provide large performance gains. As such, the network performance is improved by increasing the number of antenna arrays in a cell. The result highlights the benefit of adding DAAs in a cell. We note that it is not necessary to place a large number of antenna elements in close vicinity for reaping the benefits offered by massive MIMO. Instead, the antenna elements can be placed on separate sub-arrays. The result in Fig.~\ref{sub_arrays} indicates that we observe improvement in sum SE even when the arrays are deployed in arbitrary locations.

\begin{figure}[!t]
\centering
\includegraphics[width=21pc]{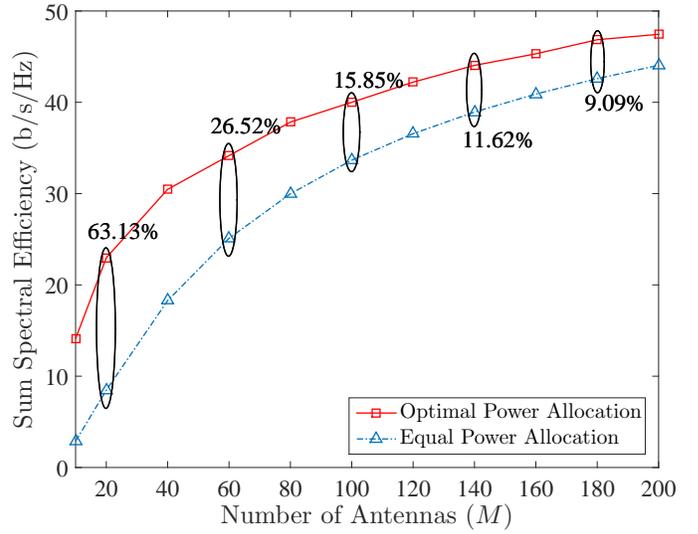}
\caption{The sum SE versus  the number of antennas per array for max-min power allocation and equal power allocation. The percentages show the improvement in sum SE provided by the max-min power allocation.}
\label{opt_power}
\end{figure}

In Fig.~\ref{opt_power}, we compare the max-min power allocation with the equal power allocation. We obtain the downlink power allocation coefficients for max-min power allocation by solving the optimization problem \eqref{opt_problem2} using CVX. For equal power allocation, we obtained the downlink power allocation coefficients using \eqref{equal_pow}. Afterwards, the sum SE for the corresponding power allocation is obtained using \eqref{SINR}. We highlight that the max-min power allocation provides a higher sum SE than the equal power allocation. For example, the max-min power allocation provides a sum SE of 34.16\,b/s/Hz compared to 25.10\,b/s/Hz provided by equal power allocation, when each array in a cell has $M=60$ antenna elements. In this case, max-min power allocation provides 26.52\% improvement in the sum SE as compared to the equal power allocation. This benefit comes in addition to the uniform user performance.

\section{Conclusion}
In this paper, we investigated the downlink power allocation in a DAA massive MIMO network. We first derived a generalized closed-form expression for the downlink SINR with correlated Rayleigh fading channels. We then formulated a max-min optimization problem based on the downlink SINR expression. We solved the max-min optimization problem and obtained the downlink power control coefficients. We then compared the performance of the optimal power allocation with equal power allocation. Our numerical results indicated that adding DAAs in the network provides a large improvement in the sum SE. Additionally, the proposed system model gives network designers great flexibility to deploy BS antenna arrays in arbitrary locations and still provide benefit from the advantages offered by massive MIMO. For future work, we will investigate heuristic power allocation algorithms and compare their performance with the max-min power allocation. Additionally, we will compare the performance of the proposed scheme with existing power allocation schemes.

\section*{Acknowledgments}
This work was supported by the Australian Government Research Training Program (RTP) Scholarship, ARC Discovery Project (DP180104062), the Swedish Research Council (VR), and ELLIIT.

\appendices
\section{Proof of Theorem \ref{theorem}}\label{SINR_proof}
The proof follows the approach in \cite{Emil2016a}. For MRT, we have $\mathbf{{a}}_{jk}^{n} = \mathbf{\widehat{h}}_{jk}^{jn}$. Accordingly, the numerator of \eqref{SINR_chan} is simplified as
\begin{align}\label{appendix1}
\mathbb{E}\left[(\mathbf{h}_{jk}^{jn})^H \mathbf{\widehat{h}}_{jk}^{jn}\right] &=\textrm{tr}(\mathbf{W}_{jk}^{n}\mathbb{E}[\mathbf{y}_{jk}^n(\mathbf{h}_{jk}^{jn})^H ]),\nonumber \\ &=
\textrm{tr}(\mathbf{W}_{jk}^{n}\mathbf{R}_{jk}^{jn}).
\end{align}
We now simplify the first term in the denominator of \eqref{SINR_chan}. We note that when $(j,k)\neq (l,i)$, $\mathbf{{h}}_{jk}^{ln}$ and $\mathbf{\widehat{h}}_{li}^{ln}$ are independent. For this case we have
\small
\begin{align}\label{first_term}
\mathbb{E}\left[\left|\sum_{n=1}^N\nu_{li}^n(\mathbf{h}_{jk}^{ln})^H \mathbf{\widehat{h}}_{li}^{ln}\right|^2\right]  & = \sum_{n=1}^N\textrm{tr}\left((\nu_{li}^n)^2\mathbf{W}_{li}^{n}\mathbf{Q}_{li}^{n}(\mathbf{W}_{li}^{n})^H\mathbf{R}_{jk}^{ln}\right).
\end{align}
\normalsize
We now consider the case where $(j,k)=(l,i)$. In this case, $\mathbf{{h}}_{jk}^{ln}$ and $\mathbf{\widehat{h}}_{li}^{ln}$ are not independent. By utilizing the fact that and $\mathbf{h}_{jk}^{ln}$ and $\mathbf{\widehat{h}}_{li}^{ln}-\mathbf{W}_{li}^{n}\mathbf{h}_{jk}^{ln}$ are independent, we have
\small
\begin{align}\label{sec_term}
\mathbb{E}\left[\left|\sum_{n=1}^N\nu_{li}^n(\mathbf{h}_{jk}^{ln})^H \mathbf{\widehat{h}}_{li}^{ln}\right|^2\right] &=  \sum_{n=1}^N\textrm{tr}\left((\nu_{li}^n)^2\mathbf{W}_{li}^{n}\mathbf{Q}_{li}^{n}(\mathbf{W}_{li}^{n})^H\mathbf{R}_{jk}^{ln}\right) \notag \\ &+ \left|\sum_{n=1}^N\textrm{tr}(\nu_{lk}^n\mathbf{W}_{li}^n\mathbf{R}_{jk}^{ln})\right|^2.
\end{align}
\normalsize
Combining \eqref{first_term} and the \eqref{sec_term}, the first term in the denominator of \eqref{SINR_chan} is written as
\small \begin{align}\label{den_first}
\mathbb{E}\left[\left|\sum_{n=1}^N\nu_{li}^n(\mathbf{h}_{jk}^{ln})^H \mathbf{\widehat{h}}_{li}^{ln}\right|^2\right]&=  \sum_{n=1}^N\textrm{tr}\left((\nu_{li}^n)^2\mathbf{W}_{li}^{n}\mathbf{Q}_{li}^{n}(\mathbf{W}_{li}^{n})^H\mathbf{R}_{jk}^{ln}\right) + \nonumber \\
&\hspace{-1.3cm}\begin{cases}
0, & (j,k)\neq(l,i)\\
\left|\sum_{n=1}^N\textrm{tr}(\nu_{lk}^n\mathbf{W}_{li}^n\mathbf{R}_{jk}^{ln})\right|^2, & (j,k)=(l,i)
\end{cases}
\end{align}
\normalsize
Substituting \eqref{appendix1} and \eqref{den_first} in \eqref{SINR_chan} we obtain \eqref{SINR}.
\section{Proof of Proposition \ref{prop_1}} \label{SOCP_proof}
We write the first constraint in the optimization problem \eqref{opt_problem1} as
\begin{align} \label{socp1}
\frac{\left|\textstyle{\sum_{n=1}^N}\nu_{jk}^n\chi_{jk}^{n}\right|^2} {\textstyle{\sum_{l,i,n}^{L,K,N}}(\nu_{li}^n)^2\zeta_{jk}^{lin} + \textstyle{\sum_{l\neq j}^{L}}|\textstyle{\sum_{n=1}^N}\nu_{lk}^n\xi_{jk}^{ln}|^2 + \sigma_{n}^2} \geq \widehat{\gamma}_{jk} .
\end{align}
Now we introduce slack variable $\nu_{lk}^n\xi_{jk}^{ln} \leq \varrho_{jk}^{lin} $ and simplify \eqref{socp1}. Accordingly, we obtain
\begin{align}\label{socp2}
   \left(\textstyle{\sum_{l,i,n}^{L,K,N}}(\nu_{li}^n)^2\zeta_{jk}^{lin} + \textstyle{\sum_{l\neq j}^{L}} (\sum_{n=1}^N\varrho_{jk}^{lin})^2 + \sigma_{n}^2\right)^{\frac{1}{2}} \leq \notag \\ \frac{1}{\sqrt{\widehat{\gamma}_{jk}}}{\left|\textstyle{\sum_{n=1}^N}\nu_{jk}^n\chi_{jk}^{n}\right|},
\end{align}
which is equivalent to
\begin{align}\label{socp3}
  \|\mathbf{x}_{jk}\| &\leq \frac{1}{\sqrt{\widehat{\gamma}_{jk}}}{\left|\textstyle{\sum_{n=1}^N}\nu_{jk}^n\chi_{jk}^{n}\right|},
\end{align}
where $\mathbf{x}_{jk} = [\mathbf{\tilde{x}}_{jk}~~\mathbf{\bar{x}}_{jk}~~\sqrt{\sigma_{n}^2}]^T$. The terms $\mathbf{\tilde{x}}_{jk}$ and $\mathbf{\bar{x}}_{jk}$ are defined in the \textit{Proposition 1}.

We highlight that the constraint given in \eqref{socp3} can be represented in the standard second-order-cone form. As such, the optimization problem in \eqref{opt_problem1} is convex. Furthermore, the constraints in \eqref{opt_problem1} are convex. Accordingly, the optimization problem in \eqref{opt_problem1} is quasi-concave. Finally, we re-write the optimization problem as \eqref{opt_problem2} given in the \textit{Proposition 1}.

\end{document}